\documentclass[10pt]{IEEEtran}
\pdfoutput=1
\newcommand{\thetitle}{Understanding Systematic Errors Through Modeling of ALMA 
Primary Beams}

\usepackage{graphicx}
\usepackage{epstopdf}

\usepackage{longtable}

\usepackage{float}

\usepackage[caption=false]{subfig}

\usepackage{amsmath,amssymb,latexsym}

\usepackage[numbers]{natbib}
\usepackage{tikz}
\usepackage{textcomp}

\newcommand\copyrighttext{%
    \footnotesize \textcopyright 2016 IEEE.  Personal use of this material is 
    permitted. Permission from IEEE must be obtained for all other uses, in any 
    current or future media, including reprinting/republishing this material 
    for advertising or promotional purposes, creating new collective works, for 
    resale or redistribution to servers or lists, or reuse of any copyrighted 
    component of this work in other works.  DOI: 10.1109 }
\newcommand\copyrightnotice{%
    \begin{tikzpicture}[remember picture, overlay]
     \node[anchor=south, yshift=10pt] at (current page.south) 
     {\fbox{\parbox{\dimexpr\textwidth-\fboxsep-\fboxrule\relax}{\copyrighttext}}};
    \end{tikzpicture}%
}

\begin{document}
\title{\thetitle}
\author{\IEEEauthorblockN{Kara Kundert}
\and
\IEEEauthorblockN{Urvashi Rau}
\and
\IEEEauthorblockN{Edwin Bergin}
\and
\IEEEauthorblockN{Sanjay Bhatnagar}
}
\maketitle
\copyrightnotice
\sloppy

\begin{abstract}
 Many aspects of the {Atacama Large Millimeter Array (ALMA)} instrument are 
 still unknown due to its young age. One such aspect is the true nature of the 
 primary beam of each baseline, and how changes to the individual primary beams 
 affect astronomical observations {when said changes are ignored during 
 imaging}.  This paper aims to create a more thorough understanding of the 
 strengths and weaknesses of ALMA through realistic modeling of the primary 
 beams and simulated observations{, which in turn can inform the user of the 
 necessity of implementing more computationally costly algorithms, such as 
A-Projection, and when simpler, quicker algorithms will suffice}.  We quantify 
our results by examining the dynamic range of each observation, along with the 
ability to reconstruct the Stokes I amplitude of the test sources.  These tests 
conclude that {for dynamic ranges of less than 1000}, for point sources and 
sources much smaller than the main lobe of the primary beam, {the accuracy of 
the primary beam model beyond the physical size of the aperture simply doesn't 
matter}. In observations of large extended sources, deconvolution errors 
dominate the reconstructed images and the individual primary beam errors were 
indistinguishable from each other.
\end{abstract}

\section{Introduction}

Radio interferometry has been a key innovation in the field of modern 
astronomy~\citep{swenson-mathur}. The ability to {synchronize} the observations 
of many separate antennas, along with dramatically improved bandwidths and 
backend technology, has led to increased observational sensitivities and 
precision in resolution. These improvements have led to the lowest systematic 
levels of noise ever seen in radio astronomy.  As the technology used to create 
the digital backends of observatories continues to see astounding rates of 
progress, new sources of systematic errors that were previously easily ignored 
are beginning to make their way above the ever falling noise floor. These newly 
unearthed sources of error have yet to be characterized and thoroughly 
understood, making them an especially treacherous threat for the astronomers 
using the observatories. This is even more prevalent in interferometric systems 
which combine numerous telescopes, often of {different} sizes, across a 
variable distance scale between antennas.

Though there are many factors contributing to each image, one that remains 
relatively unexplored is that of how the primary beam affects both the data 
collected and the final images produced~\citep{corder}. As primary beams are 
related to the baseline apertures through a simple Fourier transform, the 
primary beam of each baseline offers a unique insight into the relative state 
of the antennas in an array. In this paper, we simulate several probable 
primary beam based systematic errors that could be introduced to the data 
collected at ALMA in order to better understand the effect it could have on the 
final images produced. {These effects include but are not limited to improper 
 calibration of the secondary reflector, minor offsets in the pointing during 
 an actual observation, gravitational distortion of the aperture, and ignoring 
 parallactic angle rotation during imaging.  They also affect the final images 
 in varying ways. For example, an offset of the receiver from the focus in the 
 cryostat will change the way the aperture of the dish is illuminated, leading 
to phase errors and diminished signal amplitude}.  Finally, a pointing offset 
during observation {produces} phase errors and amplitude calibration issues 
dependent on the scale of the offset. 

\IEEEpubidadjcol

Without a thorough understanding of the health of the individual antennas and 
their relationships to each other while performing observations, new errors can 
proliferate into the images through incorrect calibration of the primary beam 
in the imaging process. Such errors include but are not limited to the hiding 
of low-amplitude astronomical sources in the side lobes of brighter sources and 
an overall increase in the noise floor of an image relative to the baseline 
thermal noise.

Intrinsically, there are two questions to be answered about these revealed 
sources of error. What effects are they imparting on astronomical observations, 
and how can those effects be corrected? The work done on this simulation aims 
primarily to answer the first question. By creating a thorough and realistic 
model of ALMA, the simulation can observe the propagation of errors generated 
by selectively introducing perturbations to the primary beams. From this 
knowledge, software can be developed to target the largest sources of error in 
data analysis.

In the following section, the mathematics of the problem are summarized. In the 
subsequent two subsections, a brief description of the ALMA instrument and the 
experimental model is provided. This is followed by the results of our testing 
and the ramifications we predict for observations on the ALMA instrument.

\section{Background}
\label{background}

In the case of the ALMA and similar instruments, the visibility function of a 
source is observed and converted into images
via a Fourier transform. As the interferometer is neither infinite in size nor 
sampling at every point in space, each visibility measurement becomes a 
discrete linear weighted sum of a range of spatial frequencies which are 
determined by the geometry of the instrument. This weighting is a function of 
the baseline apertures of the interferometer, $A$. The baseline aperture is 
found by correlating the aperture illumination functions of a pair of antennas 
in an array. This can be Fourier transformed to provide the primary beam, $B$.  
Both $A$ and $B$ determine how well a given baseline pair of antennas are able 
to see the true sky.
In attempting to recreate the true image, the estimated
primary beam of a given observation can be factored out in the final stages of
data corrections and reduction, or used to correct the image using A-projection 
(an image processing algorithm further described in Section 
\ref{future})~\citep{bhatnager}.  The better the estimated beam matches the 
true beam from the instrument,
the better the corrections on the data will be. The main question this 
simulation seeks to answer {regards} how much uncorrected changes in the 
primary beams of ALMA affect the final images produced using standard CLEAN 
tool, which is deconvolution and image correction {software} that operates by 
iteratively finding point-source peaks in an image and subtracting the scaled 
dirty beam (also called the point-spread function\footnote{Also sometimes 
referred to as the impulse response function.}, or PSF) from that 
point~\citep{hogbom}.  This method helps to minimize the effects of the 
sidelobes of the PSF and to make a true map {of the point sources' locations} 
and their amplitudes.  There is also a version of CLEAN which enables the user 
to choose model sources of multiple sizes for larger objects.  This tool is 
called {MS-CLEAN~\citep{bhatnagar2009}}.

First, the mathematical relationship between the observed visibilities and the 
sky brightness must be defined, along with how baseline apertures and primary 
beams can affect these observations. In its most basic form, for one timestep, 
baseline, and polarization, we get this Fourier pair of equations:

\begin{equation}
    \label{eq:vis}
    V_{obs} = V_{true} * A
\end{equation}
\begin{equation}
    \label{eq:im}
    I_{obs} = I_{true} \times B
\end{equation}

where $V_{obs}$ is the observed visibility, $V_{true}$ is the true visibility, 
$I_{obs}$ is the observed image, $I_{true}$ is the true image.  Note that $*$ 
denotes the convolution function, and $\times$ is multiplication.  For a source 
defined on the celestial sphere, the Fourier relationship between the 
visibility function and the image is given by the van-Cittert-Zernike theorem, 
given in symbolic form in Eq.~\eqref{eq:van-cittert}~\citep{tms}.

\begin{equation}
    \label{eq:van-cittert}
    V(u,v) \overset{\mathcal{F}}{\rightleftharpoons} I(l,m)
\end{equation}

Now let us build further complexity into equation~\eqref{eq:vis} by accounting 
for the geometry of the array, which we will mathematically describe as the 
sampling function $S(u,v)$.

\begin{equation}
    \label{eq:expanded-vis}
    V_{obs}(u,v) = [ V_{true}(u,v) * A(u,v) ] \times S(u,v)
\end{equation}

where the sampling function $S$ can be written as:

\begin{equation}
    \label{eq:sampling}
    S(u,v) = \sum_k \delta(u - u_k) \delta(v - v_k)
\end{equation}

where each $k$ represents a measurement from a single baseline.

Now finally, let us invert equation~\eqref{eq:expanded-vis} via 
Eq.~\eqref{eq:van-cittert} to yield an image of the observed source brightness 
distribution, giving us a more complex and realistic version of 
Eq.~\eqref{eq:im}.

\begin{equation}
    \label{eq:expanded-im}
    I_{obs}(l,m) = [ I_{true}(l,m) \times B(l,m) ] * I_{psf}(l,m)
\end{equation}

where $I_{psf}$ is the Fourier transform of the sampling function $S(u,v)$.

In these equations for the simplified case of a single snapshot in a single 
frequency bin and a single baseline, we find that the observed image is the 
true sky multiplied by the primary beam and then convolved with the 
point-spread function. This is the mathematical foundation of the simulation.

\subsection{The ALMA Observatory}

In the case of ALMA, the array is still in its early science phase, so some 
parameters are bound to change as the remaining components of the instrument 
gradually come online. In its final design, ALMA will consist of a 12-m array 
{composed of 50 antennas of two designs denoted as DA and DV} with baselines 
ranging from 15m to 16km, a {12-element} compact 7-m array {with baselines 
ranging from 8.5m to 30m}, and four 12-m antennas for single dish (or Total 
Power) observations, {which provide spatial information equivalent to baselines 
of 0m to 12m}.  The 7-m and total power arrays -- with their overall shorter 
baselines -- aim to fill the hole in $uv$-coverage typically seen in radio 
interferometers.  {ALMA} will be capable of observing from 31-950 GHz with full 
linear polarization (X, Y)~\citep{alma-tech-handbook}.

However, ALMA is not yet fully finished. In Early Science Cycle 
2 - which the simulation aims to model - the array has the following 
specifications: 34 12-m antennas of three separate designs, along with the 
Atacama Compact Array (ACA), which consists of nine identical 7-m 
dishes~\citep{cycle-2-capabilities}. There are also two 12-m antennas in the 
Total Power Array, used to make single dish observations in order to fill the 
central hole in the u-v place. Each antenna is equipped with receivers to 
observe at bands 3, 4, 6, 7, 8, and 9, which corresponds to wavelengths of 
about 3.1, 2.1, 1.3, 0.87, 0.74, and 0.44 mm. There are many configurations of 
the array, with maximum baselines ranging from approximately 160 m to 1.5 km, 
though the maximum baseline for bands 8 and 9 is approximately 1 km.

One fact is evidently clear - ALMA is the foremost leader in interferometric 
imaging at millimeter wavelengths. As image fidelity is most strictly limited 
by the number and coverage of samples in the u-v plane~\citep{wright}, the 
sheer number of antennas along with the variation in baseline spacings gives 
ALMA superior imaging capabilities to any other modern interferometer currently 
online. This {unprecedented} ability means that errors generated by systematic 
errors in primary beam analysis can limit the imaging capabilities of ALMA.

\begin{figure*}
    \centering
    \subfloat[Average DA Measured Aperture -- Real]{
        \includegraphics[height=2.1in]{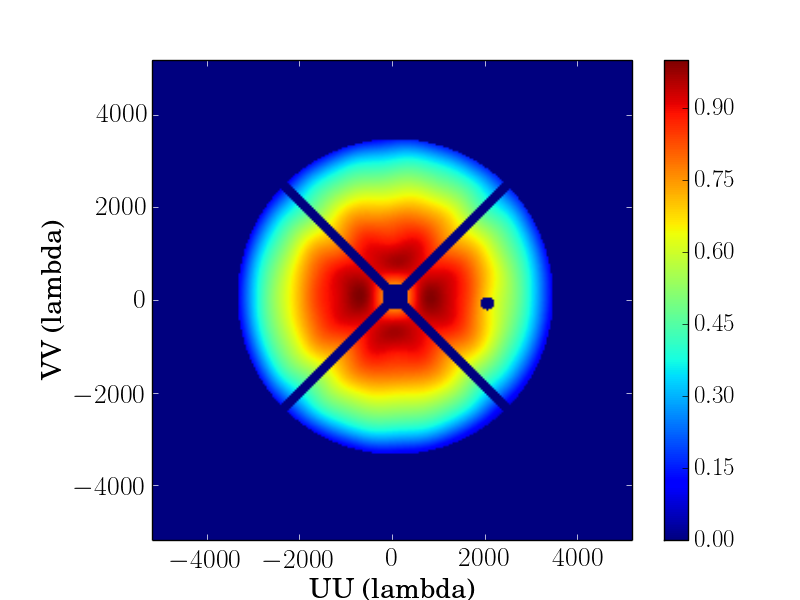}
        \label{fig:da-real}
    }
    \quad
    \subfloat[Average DA Measured Aperture -- Imaginary]{
        \includegraphics[height=2.1in]{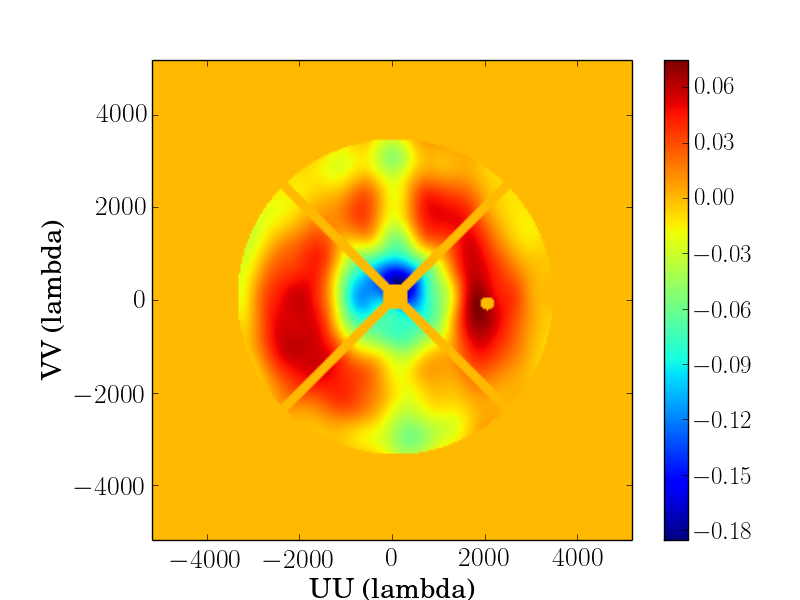}
        \label{fig:da-imag}
    }
    \quad
    \subfloat[RMS DA Measured Aperture -- Real]{
        \includegraphics[height=2.1in]{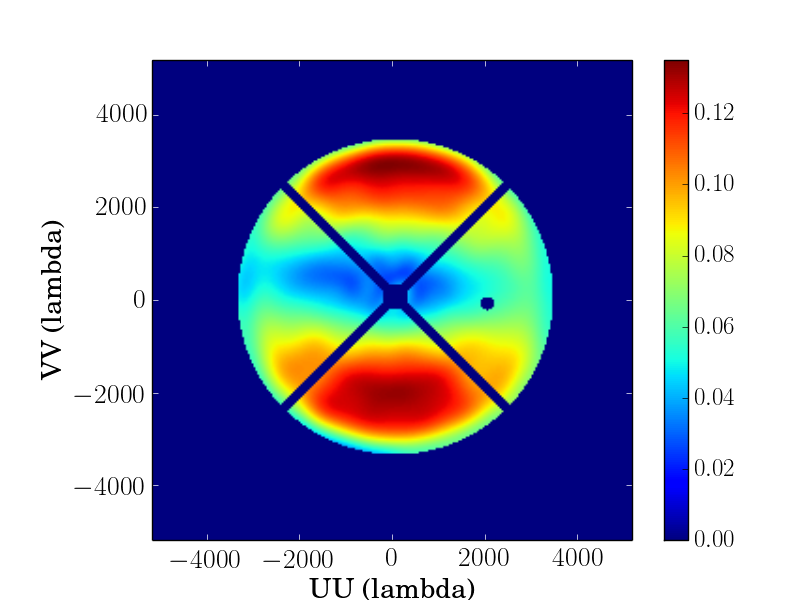}
        \label{fig:rms-da-real}
    }
    \quad
    \subfloat[RMS DA Measured Aperture -- Imaginary]{
        \includegraphics[height=2.1in]{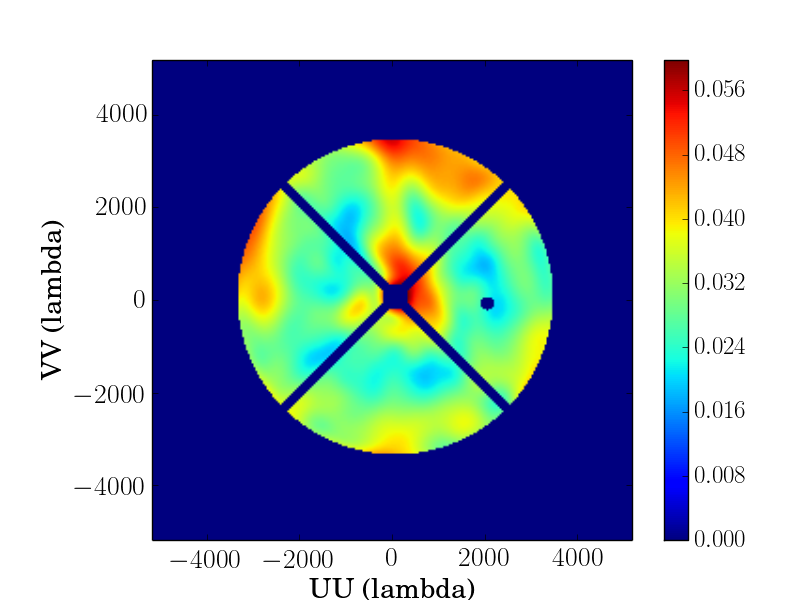}
        \label{fig:rms-da-imag}
    }
    \quad
    \subfloat[Average DV Measured Aperture -- Real]{
        \includegraphics[height=2.1in]{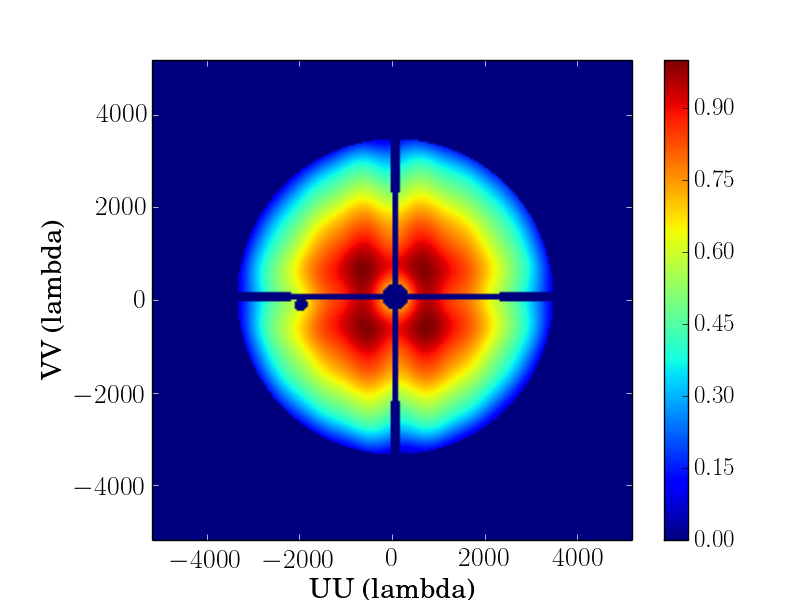}
        \label{fig:da-real}
    }
    \quad
    \subfloat[Average DV Measured Aperture -- Imaginary]{
        \includegraphics[height=2.1in]{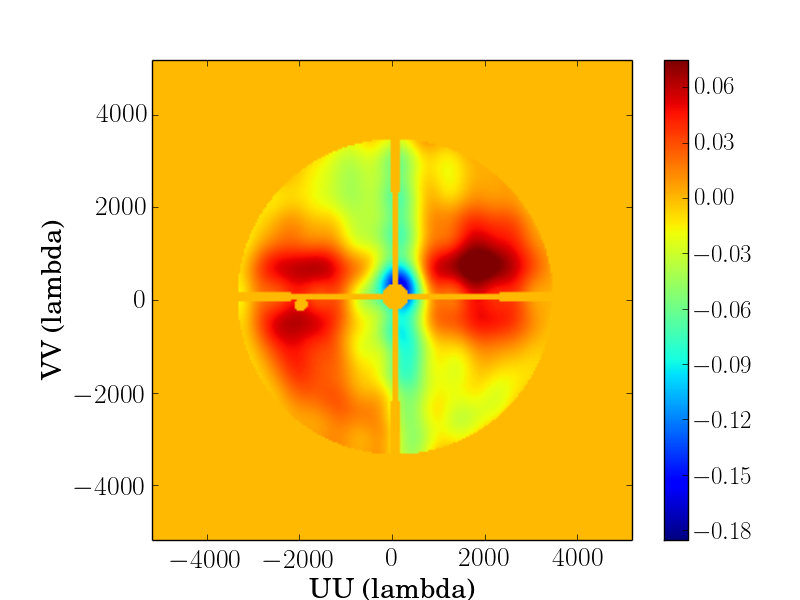}
        \label{fig:da-imag}
    }
    \quad
    \subfloat[RMS DV Measured Aperture -- Real]{
        \includegraphics[height=2.1in]{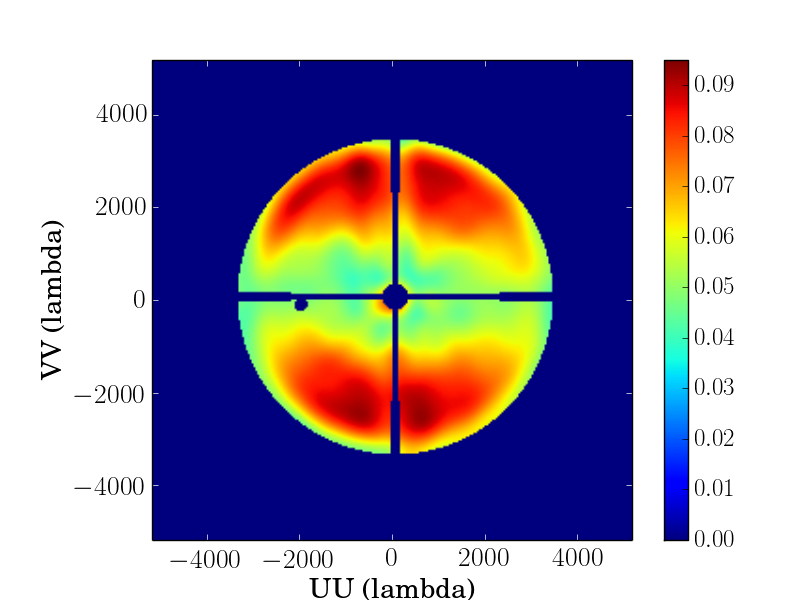}
        \label{fig:rms-da-real}
    }
    \quad
    \subfloat[RMS DV Measured Aperture -- Imaginary]{
        \includegraphics[height=2.1in]{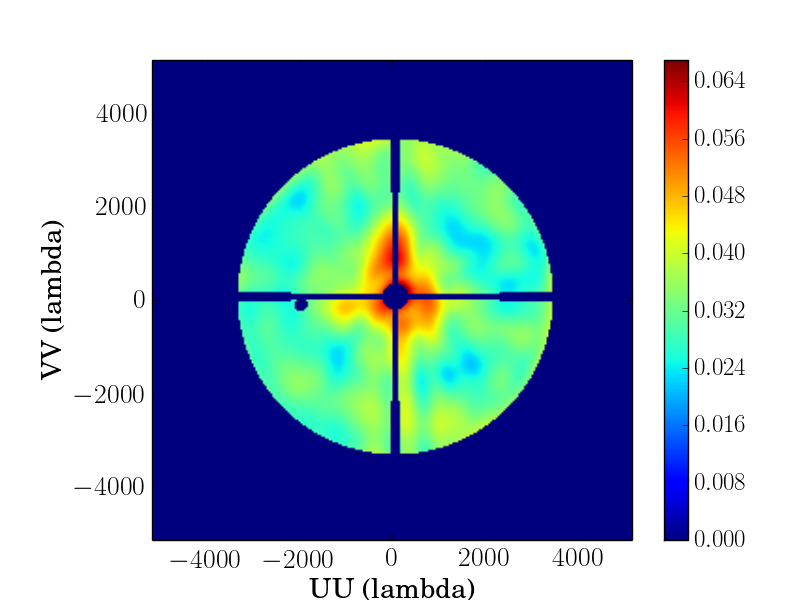}
        \label{fig:rms-da-imag}
    }
    \caption{
        {
         The average real and imaginary components of the 12m DA- and DV-type 
         apertures, pictured with the RMS variation between apertures, 
         normalized by the maximum real response. While these apertures are 
         based on true measurements of the aperture illumination, they {neglect 
         the effects of support leg diffraction}.
       }
    }
    \label{fig:apertures}
\end{figure*}

\subsection{The Model}
\label{model}

Our goal is to model a realistic interferometric array, with as many of the 
known problems of the ALMA primary beam as we know how to replicate 
computationally.  In its most basic form, the simulation does the calculations 
shown in Eq.~\eqref{eq:expanded-vis} for a set of mutually unique aperture 
models over a series of time steps. As time progresses in the simulation, so 
does the relative geometry of the array to the celestial sphere.  An individual 
simulation is composed of snapshots taken every 3 minutes over the course of 2 
hours, for a total of 40 snapshots per ``observation".  Snapshots were chosen 
over integrations to conserve computational resources, the duration was chosen 
to prevent shadowing of closely-placed antennas.

The simulation is designed to be a test of the capabilities of ALMA in Cycle 2 
of its early science testing. Therefore, the unperturbed interferometer which 
acts as the control case of the simulation has 34 identical apertures, taking 
data at 100 GHz in receiver band 3. The numerical interferometer is built by 
generating a set of antennas in an array, using the specifications on Cycle 2 
released by ALMA, including the reference location of the observatory on the 
earth, the placements of the antennas in 3D space, the Stokes parameters, the 
rest frequency of the array,  and its frequency 
resolution~\cite{cycle-2-capabilities}. In this simulation, the minimum 
baseline between antennas is approximately 15 m and the maximum is 
approximately 800 m. 

{In choosing the perturbation effects to study, two sets of simulations were 
 performed.  The first was a preliminary, exploratory version and a variety of 
 apertures, both simulated and measured, with handmade perturbations that 
 included blind and corrected pointing offsets, ellipticity, noise on the 
 aperture, parallactic angle rotation, and the uncorrected combination of the 
 DA/DV antenna types (which have a $45^\circ$ offset of the support legs).  
 Several combinations of the above effects were also tested, in order to better 
 understand how the perturbations compound with each other.  These tests 
 concluded that pointing offsets and illumination offsets were the dominant 
 source of imaging errors. A more thorough description of the layout of this 
 version and its results can be found in Appendix~\ref{prelim-tests}.  
 
 The second version of the simulation (whose results are the focus of this 
 paper) used solely measured apertures similar to the average ones shown in 
 Fig.~\ref{fig:apertures}, and focused only on the strongest perturbation 
 effects that were found in the previous round of testing. The pointing offset 
 case used identical apertures modified only by the offset itself. The 
 illumination offset case included a whole set of different apertures to test 
 the effect of the variations depicted in Fig.~\ref{fig:apertures}. Parallactic 
 angle rotation was ignored because it was found to be a relatively small 
 effect similar to combining the DA and DV antennas for this short time range 
 as can be seen in Table~\ref{tab:rms-10ants}, and because it is very 
 computationally expensive to run. Polarization was also largely ignored for 
 the sake of time.

 {Neglecting} to correct for heterogeneous antenna combinations, such as using 
 the ALMA Compact Array (ACA) with the 12-m array elements or varying designs 
 of the 12-m antennas, was also chosen as a subject for investigation in both 
 versions of the simulation. We have elected to include these results in order 
 to both validate the results of our simulation, particularly in reference to 
 previous work done on the CARMA instrument, and to understand and probe the 
 relative differences between errors due to various effects within the 
 simulation framework and improper handling of the hetereogeneous quality of 
 the array~\citep{corder}.
}

By convolving the perturbed apertures, baseline aperture functions are 
calculated for each pair of antennas, the Fourier transform of which gives the 
primary beam of that baseline. {For each baseline, the} real component of this 
primary beam is multiplied by the true sky image.  Visibilities were calculated 
from the perturbed data to produce a simulated data set using 
Eq.~\eqref{eq:expanded-vis}. The standard CASA\footnote{Common Astronomy 
Software Application} MS-CLEAN task is then run on these images of the 
``observed sky". The off-source rms-level of the image is saved to disk.  Image 
fidelity is also tested by finding the amplitude of the cleaned source divided 
by the normalized amplitude value of the CASA model primary beam at that point.  
This test was done to see how closely the primary beams need to match in order 
to get desired fidelity in image reconstruction.

\begin{figure}
    \centering
    \subfloat[Point Source]{
        \includegraphics[height=2.3in]{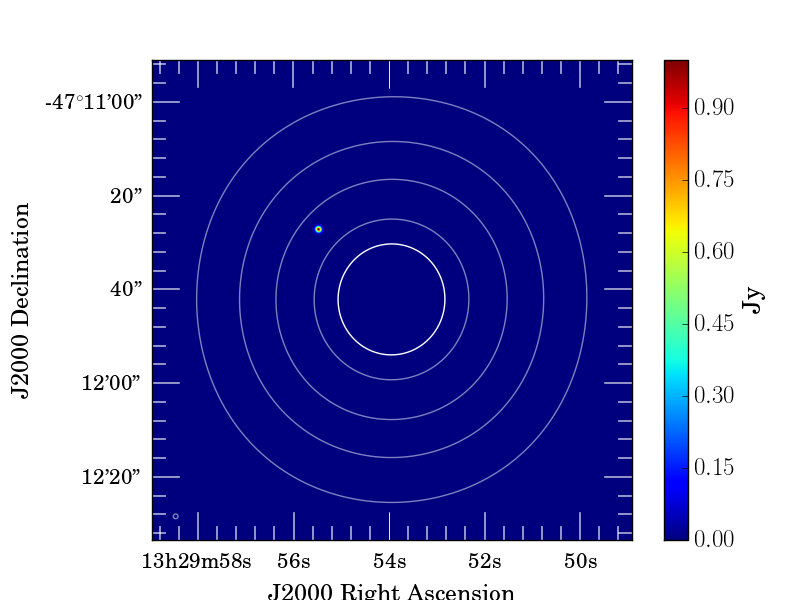}
        \label{fig:point-source}
    }
    \quad
    \subfloat[Small Extended Source]{
        \includegraphics[height=2.3in]{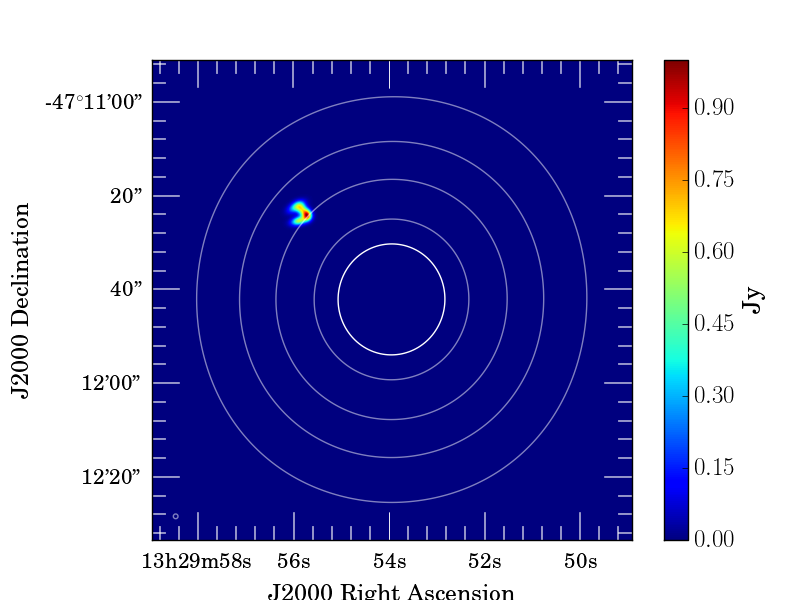}
        \label{fig:extended-source}
    }
    \quad
    \subfloat[Large Extended Source]{
        \includegraphics[height=2.3in]{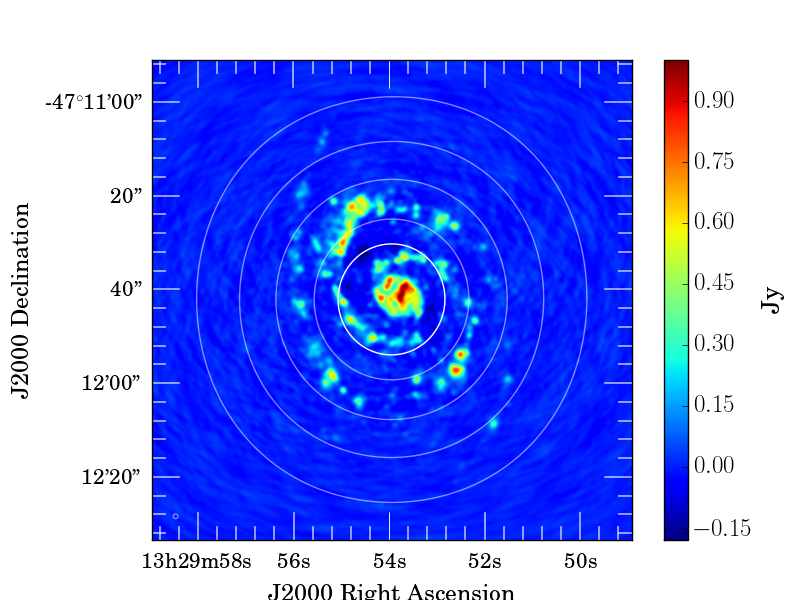}
        \label{fig:M51}
    }
    \caption{
     The simulated sources used for each test. {These are the point source 
     (Fig.~\ref{fig:point-source}), the small extended source (Fig.  
    ~\ref{fig:extended-source}), and the large extended source 
   (Fig.~\ref{fig:M51})}.  The smaller sources are offset from the center of 
   the image in order to introduce primary beam effects to the image, as a 
   small centered source will not experience primary beam related 
   perturbations.  {Also shown in each image {are contours of the main lobe of 
     the primary beam, drawn at 80\%, 60\%, 40\%, and 20\% power.}  The 
     innermost contour indicates the largest image scale that the simulation is 
    sensitive to, as set by the minimum uv-spacing.}
   } 
   \label{fig:sources}
\end{figure}

\begin{figure}
    \centering
    \subfloat[Test Antenna Locations]{
        \includegraphics[height=3.0in]{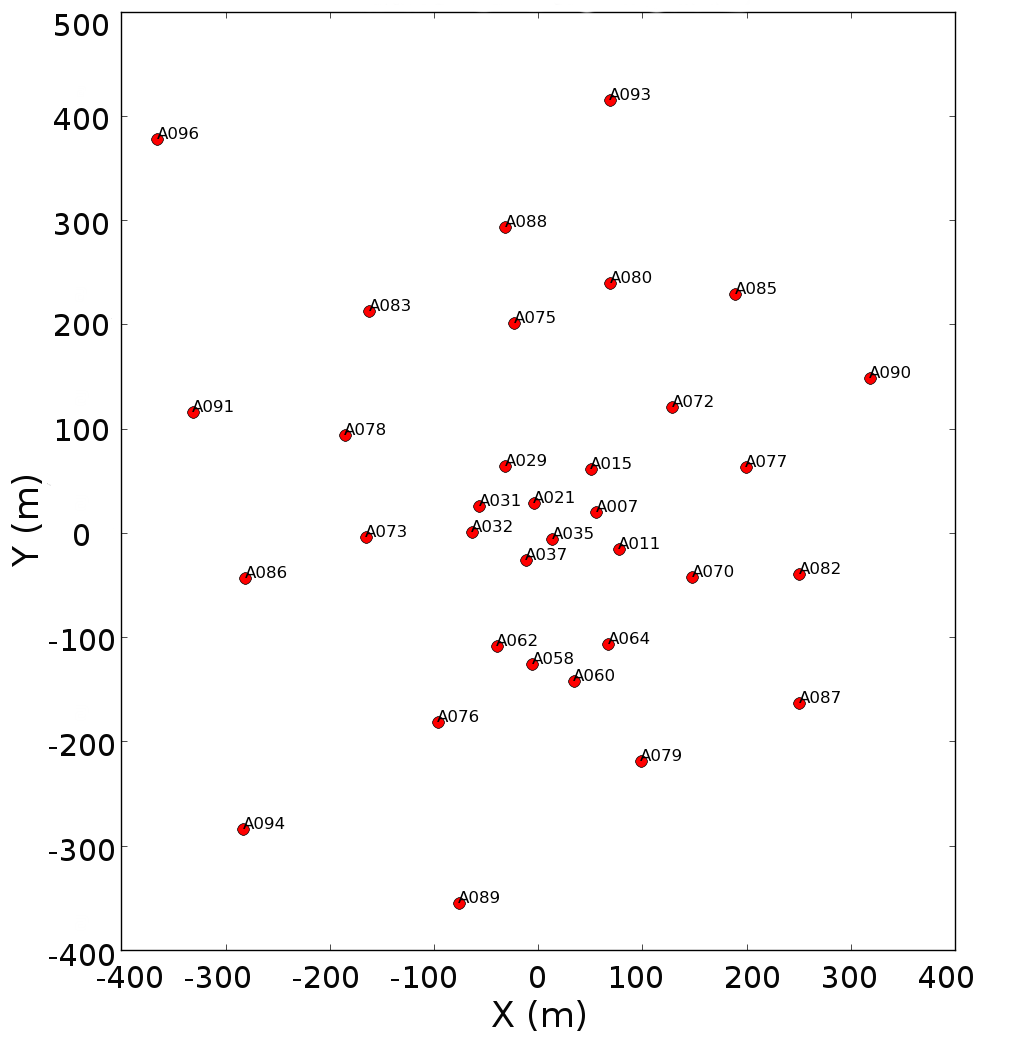}
        \label{fig:locs}
    }
    \quad
    \subfloat[Test UV-Coverage]{
        \includegraphics[height=2.75in]{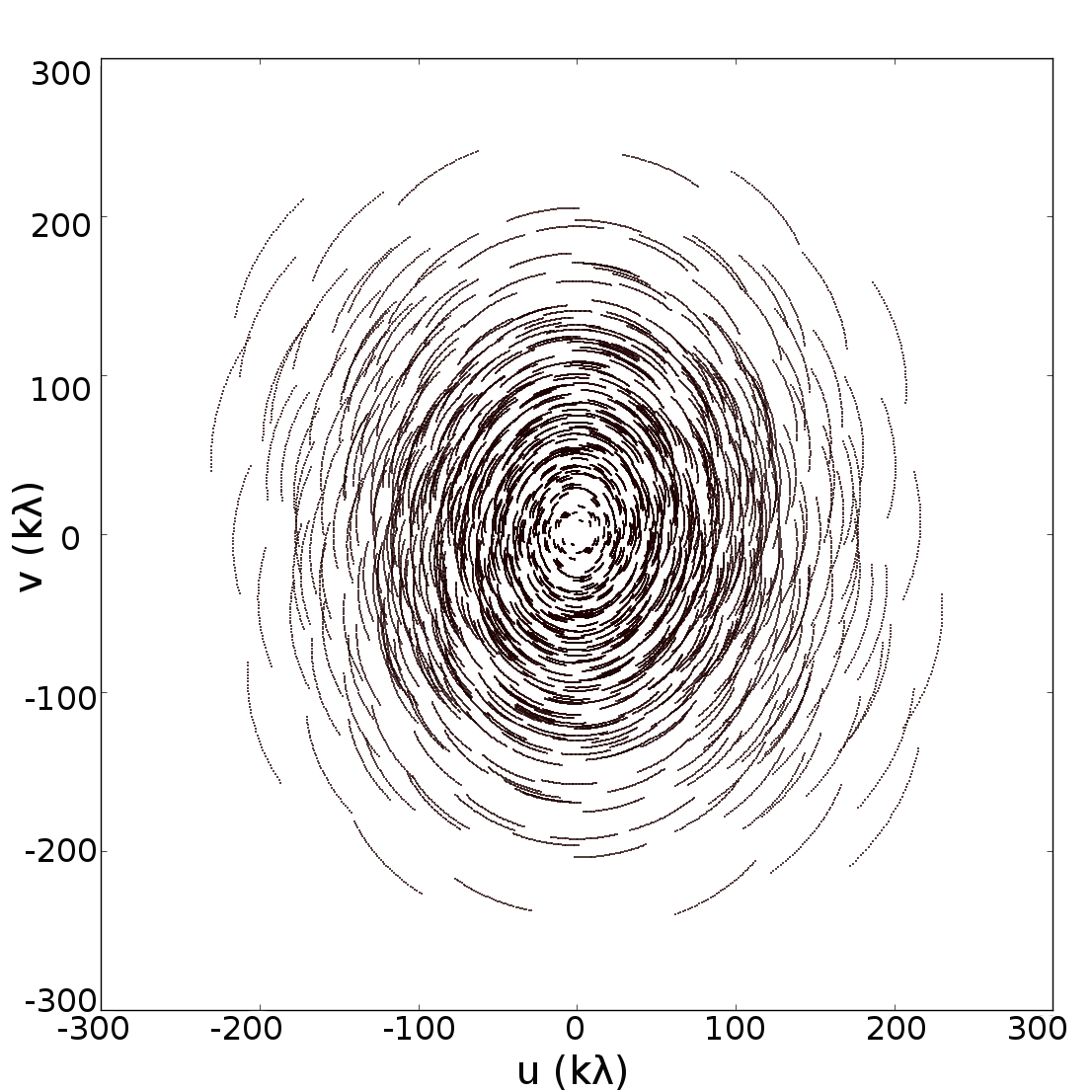}
        \label{fig:uvcoverage}
    }
    \caption{
     {The locations of the 34 antenna array used in all simulations in the 
     experiment (Fig.~\ref{fig:locs}), and the uv-coverage generated from each 
    observation (Fig.~\ref{fig:uvcoverage}).} A hole in the center of the 
    uv-field is clearly seen, as the compact array geometry was not included in 
    the simulation.
    }
    \label{fig:params}
\end{figure}

{Three main tests were run, a 1 Jy point source pointed slightly off-center, a 
 small extended source (to emulate a protoplanetary disk or small cloud), and a 
 large extended source to fill the whole primary beam, as can be seen in 
 Fig.~\ref{fig:sources}.  The source in the point source test was located at 
 approximate the 75$\%$ power point in the main lobe of an unperturbed primary 
 beam. The small extended source was centered at approximately the 60$\%$ power 
 level of the main lobe of the unperturbed primary beam. The large extended 
 source was centered.\footnote{The small sources were placed off-center in 
  order to observe the effects of primary beam corrections. A point source or 
  small source that is perfectly or nearly perfectly centered in the beam would 
  {lead to only} very small corrections from the primary beam, as it would be 
  located at nearly maximum power.  Primary beam correction errors are 
  generated in the lower-power regions and side lobes of the image.}
The {simulated} uv-coverage and antenna placement can be seen in 
Figure~\ref{fig:params}.  }

\begin{figure}
    \centering
    \includegraphics[height=2.3in]{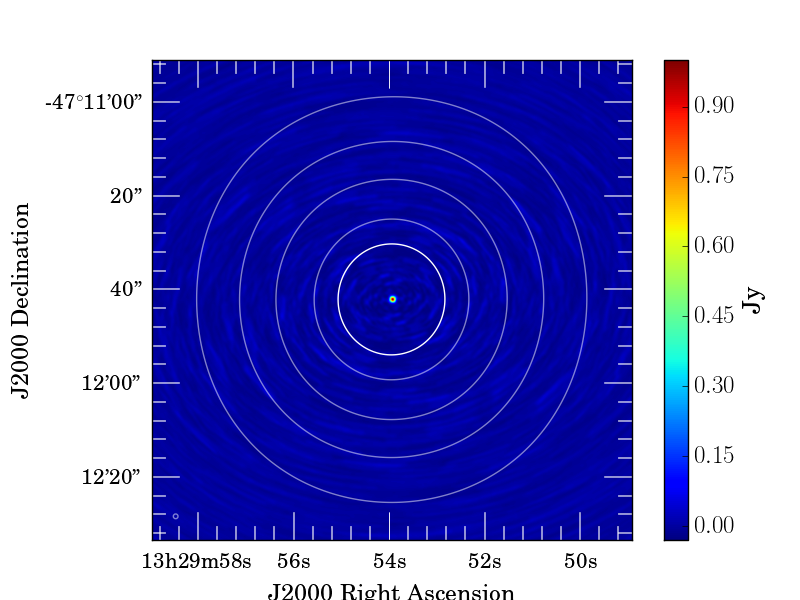}
    \caption{
        The point-spread function (PSF) of the synthesized beam used in all 
        tests, {overlaid with contours of the main lobe of the primary beam at 
         80\%, 60\%, 40\%, and 20\% power}.  Tests were run with an array of 
         34 antennas, observing 
        40 snapshots over the course of 2 hours.
    }
    \label{fig:psf}
\end{figure}

\section{Results}

{Table~\ref{tab:rms-34ants} shows that the} {most prominent primary beam} 
effect{s {in this study} were those involving} pointing offsets.  {To prevent 
 the introduction of pointing offset errors, ALMA creates pointing models for 
 each antenna in the array.  This is done in two ways. On a weekly basis, a 
 short pointing run is made, conducted with 30-50 sources spread evenly between 
 20-85$^{\circ}$ elevation.  This enables the array to ``blindly" find any 
 source in the sky with an accuracy between {2-4"} ~\citep{vanKempen}.  Beyond 
 that, in order to ensure that an observed source can be held stable at the 
 center of the beam, each scientific observation is preceded by an offset 
 pointing which enables the array to track any source to within a Gaussian RMS 
 of 0.6" of its true position~\citep{vanKempen}.  We refer to this case as a 
 ``corrected pointing offset".  While these seem like very high accuracies, 
 these numbers {are a reflection of telescope geometry, which makes them 
 frequency invariant. Given the primary beam's scaling with frequency, this 
means that the imaging effect of} pointing errors will get progressively worse 
at higher frequencies -- a corrected pointing offset at 400 GHz (or the middle 
of the range of ALMA's observing capabilities) could be just as bad as a blind 
pointing at 100 GHz, the lowest band {simulated here}. As such, the numbers 
presented for blind pointing offsets at 100 GHz can also be considered a test 
of corrected pointing offsets at 400-800 GHz, which is a relevant case to a 
typical ALMA observer.}
 
Simulated observations took place at 100 GHz, giving the primary beam a FWHM of 
approximately 45 arcseconds. At worst, the blind pointing offsets could shift 
the beam by almost $10\%$ of its width. Tests for this case gave dynamic ranges 
of approximately 1000 for both the point source and small extended source 
cases. Numbers were also given for ALMA's ability to self-correct for pointing 
offsets, which have a maximum of 0.5 arcseconds{~\citep{alma-tech-handbook}}.  
{In the best case of corrected pointing offsets for a 100 GHz observation, 
 dynamic ranges were found to be approximately 4750 for the point source case 
 and 3850 for a small extended source, or approximately $21-25\%$ of the blind 
pointing offset case.}

Illumination offsets, which occur when the optical access on the cryostat 
offsets the collected light slightly from the receiver, fell shortly below 
blind pointing offsets and slightly above the corrected pointing offset case, 
with a dynamic range near 3500 in the point source case and 2150 for the small 
extended source, {or between approximately $29-47\%$ the strength of the blind 
pointing offset case.}

Results from the small extended source case indicate generally the same trend 
in the hierarchy of perturbation of data, though at slightly higher noise 
levels in all cases.  Results from the large extended source are almost 
entirely flat, indicating that deconvolution errors are dominating over the 
problems in the primary beam.

{However, the largest error came from changing the antenna sizes, with dynamic 
 ranges from the source peak to the noise floor of around 300 in the single 
 point source test and 175 in the small extended source test. This kind of data 
 error would be generated by using the ALMA 7m and 12m arrays in conjunction 
 with each other, with the small baselines of the 7m array filling in the 
 center of the uv-coverage plane, without correcting for the use of three 
 different combinations of antenna pairs leading to three different types of 
 primary beams.  This matched the expectations set up by previous work 
 performed by Stuartt Corder on the CARMA instrument, which found that even an 
 uncorrected beam size differential of $3\%$ would result in reduction in image 
fidelity by a factor of two~\citep{corder}.}

{A set of preliminary tests took place in 2013 to determine which perturbations 
 merited more thorough investigation.  These tests investigated a much broader 
 selection of perturbation effects with less precision - the simulation used an 
 array of only 10 antennas and 4 time steps over a simulated two-hour 
 observation. Tests were run on point source observations, with one point 
 source and one multi-source test per perturbation effect. The inverse dynamic 
 ranges for these tests are given in Table~\ref{tab:rms-10ants}. 

RMS-levels for the 34 antenna test {case} can be found in 
Table~\ref{tab:rms-34ants}, which we have defined as being the noise floor of 
the simulated observation.  As the peak amplitude was normalized to 1.0 Jy, 
these numbers also indicate the inverse dynamic range in each case.  }

\begin{figure*}
    \centering
    \subfloat[No Perturbation]{
        \includegraphics[height=2.2in]{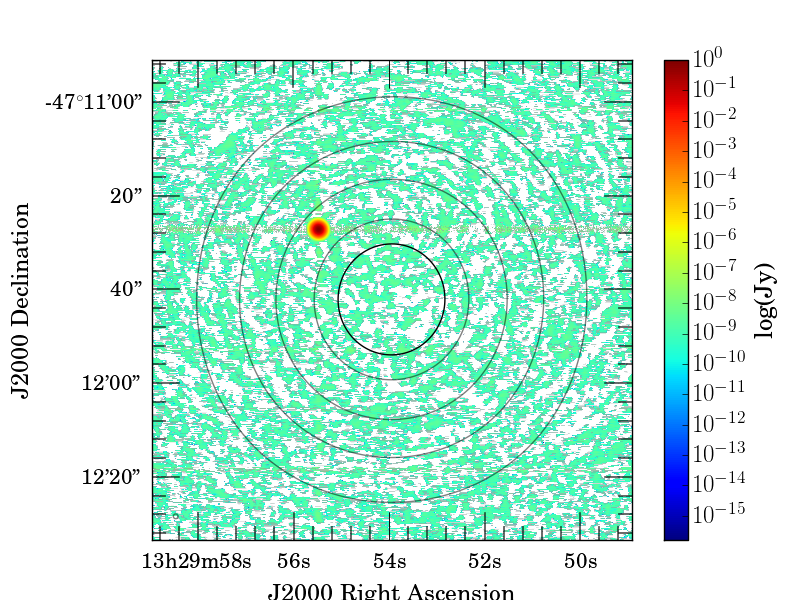}
        \label{fig:no_perturbation}
    }
    \quad
    \subfloat[Illumination Offset]{
        \includegraphics[height=2.2in]{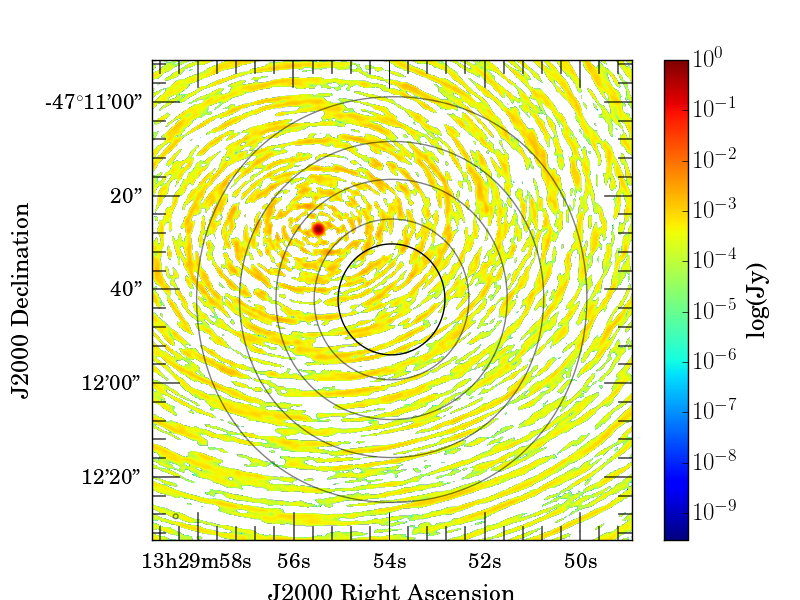}
        \label{fig:illum_offset}
    }
    \quad
    \subfloat[Corrected Pointing Offsets]{
        \includegraphics[height=2.2in]{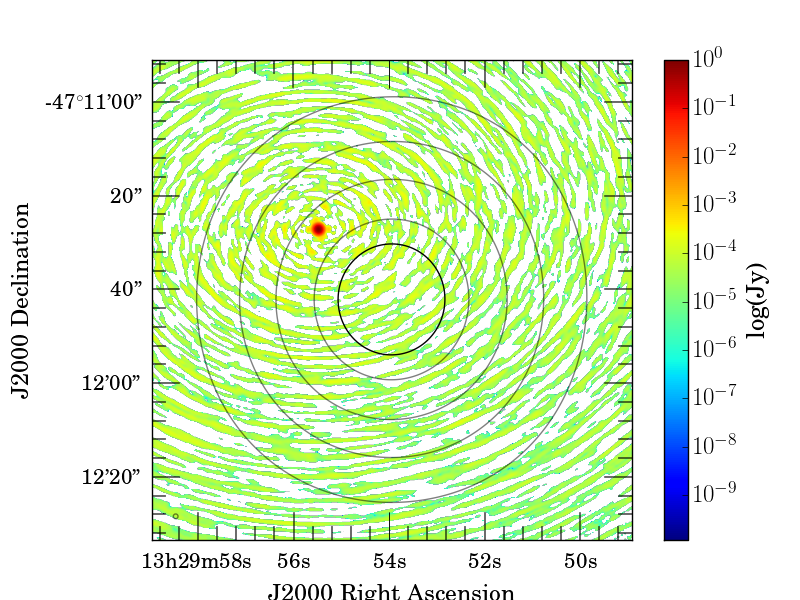}
        \label{fig:pointing_cor}
    }
    \quad
    \subfloat[Blind Pointing Offsets]{
        \includegraphics[height=2.2in]{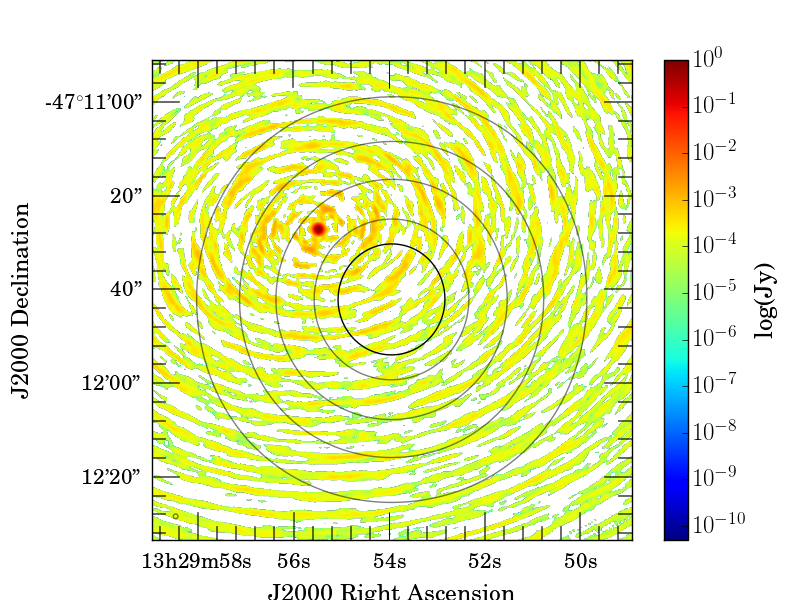}
        \label{fig:pointing}
    }
    \quad
    \subfloat[Antenna Size Difference]{
        \includegraphics[height=2.2in]{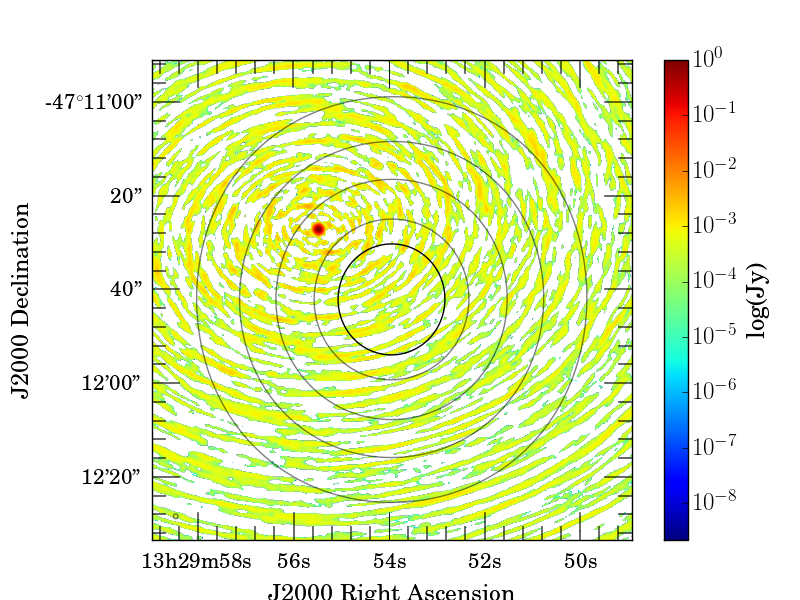}
        \label{fig:size_diff}
    }
    \quad
    \subfloat[All Effects]{
        \includegraphics[height=2.2in]{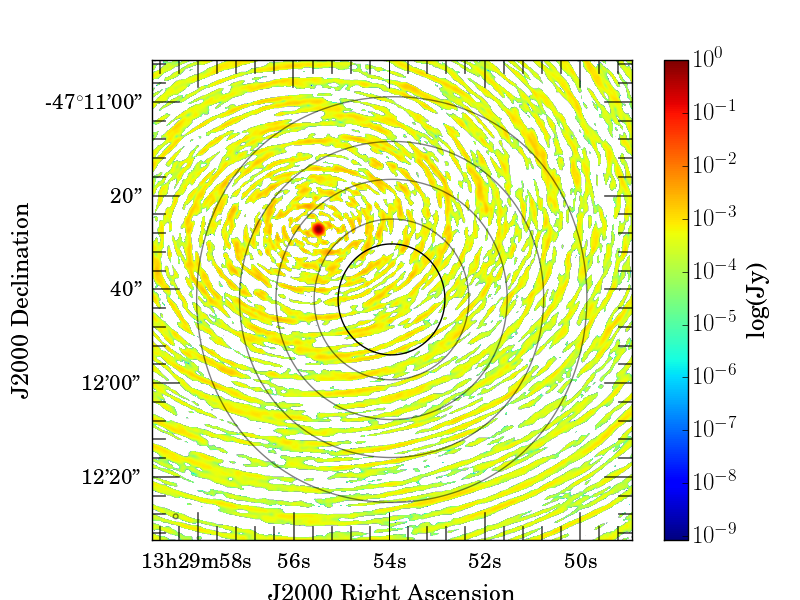}
        \label{fig:all_effects}
    }
    \caption{
        These images show the progression in overall noise as related to the 
        perturbing effects, {corresponding to successive rows in Table 
        ~\ref{tab:rms-34ants}}. All images were saturated to the same level in 
        order to make the background noise visible.
    }
    \label{fig:perturbations}
\end{figure*}

\section{Conclusions}

Initial results show that understanding the primary beam and how it affects
observations is critical to the science being done at these instruments.  {The 
strongest effect in the simulation was found to be blind pointing offsets, with 
dynamic range from source peak to noise floor limited to less than 1000 in the 
test case of a single 2 hour observation of a point source.} The next largest 
effects were illumination offsets and corrected pointing offsets, with dynamic 
ranges limited to less than 5000 in the best cases.  Preliminary testing done 
in 2013 indicated that other varieties of primary beam perturbations, e.g.  
beam rotation, ellipticity, combination of the three kinds of 12m antennas, 
were relatively minor effects, with dynamic ranges at the 10,000 level or 
higher for a single point source observation. 
 
The first effect to be concerned about would be the case of blind pointing 
offsets and high frequency corrected pointing offsets. At low frequencies, 
ensuring that all observations are done with use of pointing correction would 
certainly help. However, observations above 400 GHz are liable to see these 
same limits even with the help of the pointing guides, as pointing offsets have 
an effect that is relative to the size of the beam. As frequency goes up and 
the beam size shrinks, even a corrected pointing offset could have severe 
consequences on image quality. {Previous work from the MMA as described in ALMA 
 Memo \#95 could potentially help to mitigate the effects of pointing errors, 
 though only in the case of pointing errors that are known~\citep{holdaway}.  
 ~Since most cases of pointing offsets are random and hence difficult to 
 characterize, this algorithm will likely do little to mitigate the effects of 
 pointing offsets. However, an algorithm such as the Pointing SelfCal 
 algorithm, which attempts to solve for the unknown pointing errors and 
 incorporate those into the aperture illumination models to be used in the 
 A-Projection algorithm during imaging, could potentially drastically reduce 
 the magnitude of the errors propagated into final images through pointing 
 offsets~\citep{bhatnagar2009}. However, this algorithm has only been tested 
 for images with much lower frequencies, so it is unclear how beneficial they 
will be to the high frequency demands of ALMA.} If a method could be devised to 
correct pointing offsets at the time of observation to the 0.01 -- 
0.05 arcsecond level, even the highest frequencies would retain dynamic ranges 
of 10,000 or higher.

With all other {contributors} having much smaller effects on the residuals in 
standard imaging procedures, {effort} should be directed towards correcting for 
these larger effects before they start seriously impeding the scientific goals 
of the ALMA instrument. Efforts should also be made to improve deconvolution 
algorithms, in order to improve imaging of extended objects to the point that 
primary beam perturbations can begin to affect the data.

{Finally, the {lack of proper calibration and clean-up of }mixed-array data was 
 found to be a disastrous source of excess image noise, limiting the image 
 dynamic range to less than 
 300 in the best case of an unresolved point source over one 2-hour 
 observation. This was the expected result which affirmed the accuracy of our 
 ALMA model relative to previous similar tests done at CARMA~\citep{corder}.  
 The 7m and 12m arrays are designed to be used together, with the 7m array 
 filling in the uv-coverage hole that couldn't be filled with the larger 12m 
 dishes. If the algorithms and software used for full-beam imaging {do not} 
 account for the valid physical differences in the apertures (and hence their 
 primary beams), our testing indicates a starkly limited dynamic range on 
observations.   This kind of algorithm is already available in the {ALMA 
Science Pipeline}, and therefore shouldn't affect further imaging efforts.}

{
\subsection{Notes for ALMA Users}

These tests aimed to quantify the dynamic ranges at which different errors 
begin to affect the quality of ALMA images when left ignored during the imaging 
process. This information is useful to decide when one needs to invoke 
computationally expensive algorithms, such as A-Projection, when simpler 
methods and algorithms {will clean data} without sacrificing image quality.  

Taking the full complexity of ALMA beams into account via A-Projection is 
computationally expensive because with N different apertures, {there are} 
N(N-1)/2 separate primary beams (one per unique baseline), each of which has to 
be precomputed and cached at different parallactic angles.  {Taking} the 
example of the final, full ALMA array (with 66 antennas) and sky rotations 
calculated at $5^{\circ}$ increments, over 25,000 separate convolution 
functions would have to calculated over a 4 hour observation.  Computing one 
baseline beam and rotating it incurs errors, so each one must be computed 
separately. The $uv$-resolution at which this must be calculated also depends 
on the particular setup of an individual imaging run. All in all, it is clear 
that it is of immense practical value to understand precisely when this full 
detail is needed and when various approximations will suffice.

Our results {provide the following guidelines:} for dynamic ranges of less than 
1000, the accuracy of the primary beam model (beyond physical aperture size) 
simply does not matter.  At dynamic ranges beyond 1000, pointing offsets will 
begin to affect final image quality, followed by changes in beam shape due to 
illumination offsets beyond dynamic ranges of 5000. Finally, parallactic angle 
rotation and the beam differences between the DA/DV antennas will {become 
appreciable} at dynamic ranges beyond 10,000.

}

\section{Future Work}
\label{future}

Future tests could investigate the benefits of A-projection in image 
correction{~\citep{bhatnager}}.  A-projection uses a model of the aperture in 
the initial steps of imaging and deconvolution in order to remove its effects 
from the final image.  The idea goes that the baseline apertures $A$ have 
contributed various effects into the visibilities as they are observed, as 
described in~\eqref{eq:vis}. In order to remove those effects, the observed 
visibilities must be convolved with the inverse of the baseline aperture 
function $A^{-1}$, as seen in~\eqref{eq:a-proj}. The baseline aperture and its 
inverse will then form a unity matrix, returning the true visibilities, which 
can be Fourier transformed into the true image.

\begin{equation}
    \label{eq:a-proj}
    V_{true} = A^{-1} * V_{obs}
\end{equation}
\begin{equation}
    \label{eq:a-inv}
    A^{-1} = \frac{A^{\dagger}}{AA^{\dagger}}
\end{equation}

This process takes place during the regridding of the visibilities, so that 
$A^{-1}$ replaces the weighting function that evenly samples the u-v plane.  
However, $AA^{\dagger}$ may contain zero valued components. As such, formal 
division by $AA^{\dagger}$ is an extremely dangerous computation to make, as it 
has the potential to fill the u-v plane with divide-by-zero errors and create 
an entirely erroneous image. Instead, the Fourier transform relationship 
between the aperture and the primary beam is used to rewrite~\eqref{eq:a-proj} 
and \eqref{eq:a-inv} as

\begin{equation}
    \label{eq:a-proj-cor}
    V_{cor} = A^{\dagger} * V_{obs}
\end{equation}
\begin{equation}
    \label{eq:imcor}
    B^2 \overset{\mathcal{F}}{\rightleftharpoons} AA^{\dagger}
\end{equation}

$B^2$ is later factored out of the final image to recover $I_{true}$, 
minimizing the divide-by-zero errors while maintaining the integrity of the 
A-projection relationship. This method allows for a great deal of instrumental 
image correction prior to actually entering the image domain, assuming an 
accurate model of the aperture functions $A$ is used.

Another subject for further exploration would be the utilization of the full 
complex beam in imaging in order to better understand the propagation of 
primary beam errors through the Stokes polarization planes, {such as the effect 
of beam squint, which would appear in the Stokes Q plane in ALMA data}.  While 
our simulation used the full complex apertures to construct complex beams, only 
the real component of the primary beam was used in imaging as we were focusing 
solely on the Stokes I plane.  {Furthermore, the complicated complex structure 
 of the aperture functions and their significant antenna-to-antenna variations 
 (as seen in the real and imaginary RMS apertures presented in 
 Fig.~\ref{fig:apertures}) will lead to residual phase structure in the complex 
 primary beam which will affect Stokes I imaging as well.  

 Since the computational cost of A-Projection increases significantly if 
 antenna-to-antenna variation is included, it is of interest to determine 
 tolerance levels at which, if correctable in the hardware, antenna-to-antenna 
 variations may be ignored up to some high dynamic range. In the results 
 presented in this study, the effect of aperture illumination offsets on the 
 real part of the primary beam is already the most significant factor beyond 
 uncorrected pointing offsets and neglecting to correct for heterogeneous array 
 elements.  The effects of aperture illumination are likely to become more 
 significant when the residual phase structures in the beams are also included.
Further studies will evaluate the significance of this effect.} 

Finally, a version of this simulation performed with full high-resolution 
measured apertures, {including} antenna arm diffraction, could also provide 
helpful information to ALMA users. So far no such testing has taken place for 
lack of such apertures.

\section{Acknowledgements}

{It brings the authors great pleasure to thank Ken Kundert, Rick Perley, and 
Tom Rice for useful discussions. This work was supported in part by the 
National Radio Astronomy Observatory's Summer Student Research Assistantship 
Program.}

\bibliographystyle{IEEEtranN}
\bibliography{IEEEabrv,primary-beams}{}

\begin{thebibliography}{11}
\providecommand{\natexlab}[1]{#1}
\providecommand{\url}[1]{#1}
\csname url@samestyle\endcsname
\providecommand{\newblock}{\relax}
\providecommand{\bibinfo}[2]{#2}
\providecommand{\BIBentrySTDinterwordspacing}{\spaceskip=0pt\relax}
\providecommand{\BIBentryALTinterwordstretchfactor}{4}
\providecommand{\BIBentryALTinterwordspacing}{\spaceskip=\fontdimen2\font plus
\BIBentryALTinterwordstretchfactor\fontdimen3\font minus
  \fontdimen4\font\relax}
\providecommand{\BIBforeignlanguage}[2]{{%
\expandafter\ifx\csname l@#1\endcsname\relax
\typeout{** WARNING: IEEEtranN.bst: No hyphenation pattern has been}%
\typeout{** loaded for the language `#1'. Using the pattern for}%
\typeout{** the default language instead.}%
\else
\language=\csname l@#1\endcsname
\fi
#2}}
\providecommand{\BIBdecl}{\relax}
\BIBdecl

\bibitem[{Swenson} and {Mathur}(1968)]{swenson-mathur}
G.~W. {Swenson}, Jr. and N.~C. {Mathur}, ``The interferometer in radio
  astronomy,'' in \emph{Proceedings of the IEEE}, vol.~56, December 1968.

\bibitem[{Corder}(2009)]{corder}
S.~A. {Corder}, ``Optimizing image fidelity with arrays,'' Ph.D. dissertation,
  California Institute of Technology, 2009.

\bibitem[{Bhatnagar} et~al.(2008){Bhatnagar}, {Cornwell}, {Golap}, and
  {Uson}]{bhatnager}
S.~{Bhatnagar}, T.~J. {Cornwell}, K.~{Golap}, and J.~M. {Uson}, ``Correcting
  direction-dependent gains in the deconvolution of radio interferometric
  images,'' \emph{Astronomy \& Astrophysics}, vol. 487, no.~1, pp. 419--429,
  2008.

\bibitem[{H\"{o}gbom}(1974)]{hogbom}
J.~A. {H\"{o}gbom}, ``Aperture synthesis with a non-regular distribution of
  interferometer baselines,'' \emph{Astronomy \& Astrophysics Supplement},
  vol.~15, pp. 417--426, 1974.

\bibitem[{Bhatnagar}(2009)]{bhatnagar2009}
S.~{Bhatnagar}, ``Calibration and imaging challenges at low radio frequencies:
  An overview of the state of the art,'' in \emph{The Low-Frequency Radio
  Universe}, ser. ASP Conference Series, vol. LFRU, June 2009.

\bibitem[{Thompson} et~al.(1986){Thompson}, {Moran}, and {Swenson}]{tms}
A.~R. {Thompson}, J.~M. {Moran}, and G.~M. {Swenson}, Jr., \emph{Interferometry
  and Synthesis in Radio Astronomy}.\hskip 1em plus 0.5em minus 0.4em\relax
  John Wiley \& Sons, Inc., 1986.

\bibitem[Lundgren(2013)]{alma-tech-handbook}
A.~Lundgren, \emph{ALMA Cycle 2 Technical Handbook}, 1st~ed., ALMA, 2013.

\bibitem[ALMA(2013)]{cycle-2-capabilities}
ALMA, \emph{Cycle 2 Capabilities}, ALMA, 2013.

\bibitem[{Wright}(1999)]{wright}
M.~C.~H. {Wright}, ``Image fidelity,'' Berkeley Illinois Maryland Association,
  Memo~73, 1999.

\bibitem[{van Kempen} et~al.(2012){van Kempen}, {Corder}, {Lucas}, and
  {Mauersberger}]{vanKempen}
T.~{van Kempen}, S.~{Corder}, R.~{Lucas}, and R.~{Mauersberger}, ``How alma is
  calibrated: I. antenna-based pointing, focus and amplitude calibration,'' in
  \emph{ALMA Newsletter}, ser. ALMA in-depth.\hskip 1em plus 0.5em minus
  0.4em\relax ALMA Observatory, February 2012, no.~9, pp. 8--16.

\bibitem[{Holdaway}(1993)]{holdaway}
M.~A. {Holdaway}, ``Imaging with known pointing errors,'' Millimeter Array,
  Memo~95, 1993.

\end{thebibliography}

\onecolumn
\appendices

\section{Preliminary Results}
\label{prelim-tests}

Preliminary tests were performed over the summer and fall of 2013, focusing 
mainly on building a functioning architecture upon which further, more 
realistic tests could later be performed. As such, the main structure of the 
model is the same as described in Section~\ref{model}: a numerical array is 
constructed by generating a set of apertures in an array, those apertures are 
converted into primary beams based on their location and perturbation 
information, these primary beams are multiplied by the true sky image to create 
an ``observed sky", and CLEAN is run on these images. Statistics are then 
measured off of these images, such as the rms-noise level, dynamic range of the 
peak-to-noise, and the image fidelity.  

In this preliminary testing period, the parameters for testing were chosen in 
an effort to minimize testing time while still getting useful results {for a 
wide range of aperture types and perturbation effects}.  As such, the 
simulation was run with 10 antennas taking 4 snapshots over a two-hour 
``observation". Two main tests were run, a single point source test and a 
multi-source test with four point sources. The sources were places at various 
points in the main lobe of the primary beam (75\%, 60\%, and 30\% power points 
of the PB), with the fourth source located in the first side lobe.  The source 
in the single source case was located at approximately the 85\% power point of 
the main lobe. This was done in an attempt to investigate how the side lobes of 
the primary beam change with perturbations, the effects of which are expected 
to be much different than in the main lobe.

This overall computational minimization allowed us to factor in full 
time-dependence, {including parallactic angle rotation}.  We also tested a 
variety of types of apertures of varying scales of realism.  In the most basic 
test, there was a script in the simulation to generate a {Numeric Python 
(NumPy)} array containing the array.  In this case, the array was completely 
flat and real, and contained very little of the more interesting and realistic 
structures of the other tested apertures.  Tests were also run using two kinds 
of ray-traced aperture models - one made in CASA using the ALMA parameters, and 
one made by Dr.  Stuartt Corder {with an imposed feed-leg shadow mask excluding 
diffraction effects} through data taken in holography measurements.
While both the CASA ray-traced apertures and the measured apertures are clearly 
much more realistic than the Python-generated apertures, only the measured 
apertures include imaginary structure, making them the only ones to enable 
Stokes polarization analysis\footnote{Note that while there was some initial 
 investigation done of primary beam effects and polarization, this was 
 eventually set aside before the version of the simulation detailed in this 
 paper. As such, only the real component of the beam was used in image 
calculations.}. 

Below is a table displaying the inverse dynamic ranges for each of the tests 
performed, where the inverse dynamic range is defined as the rms-noise power 
level of the image divided by the peak source power level. We tested the 
effects of size differences in the apertures (denoted as 7m, 12m in the table), 
parallactic angle rotation of the beams, change in the orientation of the beams 
(to model the difference between the DA/{(DV/PM)\footnote{The DV and PM models 
  of antenna are very similar to each other in design, so any differences in 
  their primary beam are expected to have essentially undetectable effects on 
imaging. However, the PM model antenna is only used in the Total Power Array.}}
  antenna types in ALMA, one of which has support legs that have been rotated 
  by $45^\circ$ relative to the other), blind and corrected pointing offsets, 
  and illumination offsets of the beams. We also tested various combinations of 
  these effects, which can be found in the tables below. We were only able to 
  test size difference effects using the NumPy-generated beams as we hadn't 
  been able to acquire measured or CASA-ray traced apertures for the 7m 
  antennas. We were also only able to perform illumination offset tests on the 
  measured beams, as they were the only apertures that had illumination 
  information.

\begin{table}[H]
    \caption{
        Approximate inverse dynamic range levels found in the corrected images 
        of the preliminary runs of the simulation, using 10 antennas and 4 time 
        steps.
    }
    \label{tab:rms-10ants}
    \centering
    \begin{tabular}{|p{8cm}|c|c|}
    \hline
    & Single Source & Multi-Source  \\
    \hline
    7m, 12m, NumPy-Generated & $4.55 \times 10^{-5}$ & $2.86 \times 10^{-4}$ \\
    \hline
    Unperturbed, NumPy-Generated & $3.17 \times 10^{-9}$ & $2.99 \times
    10^{-9}$ \\
    \hline
    Parallactic Angle Rotation, NumPy-Generated & $6.74 \times 10^{-7}$ &
    $1.84 \times 10^{-5}$ \\
    \hline
    DA/DV, NumPy-Generated & $3.91 \times 10^{-6}$ & $3.89 \times 10^{-5}$ \\
    \hline
    DA/DV with Parallactic Angle Rotation, NumPy-Generated & $3.99 \times
    10^{-6}$ & $4.23 \times 10^{-5}$ \\
    \hline
    Blind Pointing Offsets, NumPy-Generated & $2.85 \times 10^{-5}$ &
    $6.18 \times 10^{-5}$ \\
    \hline
    Corrected Pointing Offsets, NumPy-Generated & $7.82 \times 10^{-6}$ &
    $1.69 \times 10^{-5}$ \\
    \hline
    Corrected Pointing Offsets with Parallactic Angle Rotation,
    NumPy-Generated & $4.78 \times 10^{-6}$ & $2.74 \times 10^{-5}$ \\
    \hline
    DA/DV with Corrected Pointing Offsets, NumPy-Generated & $5.39 \times
    10^{-6}$ & $5.84 \times 10^{-4}$ \\
    \hline
    Unperturbed, CASA Ray-Traced& $3.19 \times 10^{-9}$ & $3.60 \times
    10^{-9}$ \\
    \hline
    Parallactic Angle Rotation, CASA Ray-Traced & $8.52 \times 10^{-7}$ &
    $1.32 \times 10^{-5}$ \\
    \hline
    DA/DV, CASA Ray-Traced & $3.42 \times 10^{-7}$ & $4.53 \times 10^{-6}$ \\
    \hline
    DA/DV with Parallactic Angle Rotation, CASA Ray-Traced & $8.52 \times
    10^{-7}$ & $1.32 \times 10^{-5}$ \\
    \hline
    Blind Pointing Offsets, CASA Ray-Traced & $2.29 \times 10^{-5}$ &
    $4.81 \times 10^{-5}$ \\
    \hline
    Corrected Pointing Offsets, CASA Ray-Traced & $6.19 \times 10^{-6}$ &
    $1.98 \times 10^{-5}$ \\
    \hline
    Corrected Pointing Offsets with Parallactic Angle Rotation, CASA
    Ray-Traced & $7.28 \times 10^{-6}$ & $2.20 \times 10^{-5}$ \\
    \hline
    DA/DV with Corrected Pointing Offsets, CASA Ray-Traced & $5.32 \times
    10^{-6}$ & $1.54 \times 10^{-5}$ \\
    \hline
    Unperturbed, Measured & $4.47 \times 10^{-9}$ & $7.76 \times
    10^{-9}$ \\
    \hline
    Parallactic Angle Rotation, Measured & $4.05 \times 10^{-6}$ &
    $2.48 \times 10^{-5}$\\
    \hline
    DA/DV, Measured & $3.13 \times 10^{-6}$ & $2.84 \times 10^{-5}$ \\
    \hline
    DA/DV with Parallactic Angle Rotation, Measured & $5.83 \times
    10^{-6}$ & $2.78 \times 10^{-5}$ \\
    \hline
    Blind Pointing Offsets, Measured & $1.27 \times 10^{-5}$
    & $5.90 \times 10^{-5}$ \\
    \hline
    Corrected Pointing Offsets, Measured & $4.70 \times 10^{-6}$ &
    $1.33 \times 10^{-5}$\\
    \hline
    Corrected Pointing Offsets with Parallactic Angle Rotation, Measured & 
    $8.78 \times 10^{-6}$ & $2.93 \times 10^{-5}$ \\
    \hline
    DA/DV with Corrected Pointing Offsets, Measured & $2.49 \times
    10^{-5}$ & $6.28 \times 10^{-5}$ \\
    \hline
    Illumination Offsets, Measured & $8.85 \times
    10^{-5}$ & $2.46 \times 10^{-5}$ \\
    \hline
    \end{tabular}
\end{table}

\section{Full Model Results}

\begin{table}[H]
    \caption{
         Approximate inverse dynamic range levels found in the corrected images 
         of the full simulation, using 34 antennas and 40 time steps. All tests 
         used measured apertures. The ``all effects" case includes corrected 
         pointing offsets, illumination offsets, and size difference 
         perturbations in the primary beams used in the imaging process.
    }
    \label{tab:rms-34ants}
    \centering
    \begin{tabular}{|p{3.5cm}|c|c|c|}
    \hline
    & Point Source & Small Extended Source \linebreak (Off-Center) & Large 
    Extended Source \\
    \hline
    No Perturbation & $6.0 \times 10^{-8}$ & $7.6 \times 10^{-5}$ & 0.013 \\
    \hline
    Corrected Pointing Offsets & $2.1 \times 10^{-4}$ & $2.6 \times 10^{-4}$ & 
    0.013 \\
    \hline
    Illumination Offsets & $2.8 \times 10^{-4}$ & $4.6 \times 10^{-4}$ & 0.013 
    \\
    \hline
    Blind Pointing Offsets & $1.0 \times 10^{-3}$ & $9.6 \times 10^{-4}$ & 
    0.013\\
    \hline
    Size Difference & $3.3 \times 10^{-3}$ & $5.7 \times 10^{-3}$ & 0.014 \\
    \hline
    All Effects & $3.5 \times 10^{-3}$ & $6.1 \times 10^{-3}$ & 
    0.014 \\
    \hline
    \end{tabular}
\end{table}

\end{document}